\documentclass{llncs}

\usepackage{enumitem}
\usepackage{algorithm}
\usepackage{algpseudocode} 
\usepackage{threeparttable}
\usepackage{amsmath}
\usepackage{amsfonts}
\usepackage{amssymb}
\usepackage{wrapfig}
\usepackage{graphicx}
\usepackage[usenames,dvipsnames,svgnames,table]{xcolor}
\usepackage{comment}
\usepackage{url} 
\usepackage{multirow}

\newcommand{\G}{\textbf{G}\xspace}
\newcommand{\F}{\textbf{F}\xspace}

\newcommand{\U}{\textbf{U}\xspace}
\newcommand{\sig}[1]{{\small\textsf{#1}\xspace}}

\newcommand{\port}[1]{{\small\textsf{#1}\xspace}}

\usepackage{xspace}
\newcommand{\comments}[1]{}

\newcommand{\phio}{\textit{Robot}}
\newcommand{\buffer}{\textit{Buffer}}

\newcommand{\tEn}[1]{\textit{en}^t({#1})}

\newcommand{\stay}[1]{\delta_{#1}}

\newcommand{\ci}{\textit{CI}}

\newcommand{\efsmt}{{\small\textsf{EFSMT}}\xspace}
\newcommand{\esolver}{{\small\textsf{E-solver}}\xspace}
\newcommand{\fsolver}{{\small\textsf{F-solver}}\xspace}

\newcommand{\obs}[1]{\textcolor{OrangeRed}{OBS: #1}}

\newcommand{\mycomment}[1]{}
 
\newcommand{\deadd}{\phi_{deadlock}} 
\newcommand{\dead}{\phi_{deadlock}(\vec{v})}
\newcommand{\syst}{\mathcal{S}}
\newcommand{\systik}{\mathcal{S}_{inv, k}}
\newcommand{\mon}[1]{C_{\neg\phi_{int},#1}}

\newcommand{\pintt}{\phi_{int}}
\newcommand{\isyst}{\phi_{\syst}}
\newcommand{\Reach}{\mathit Reach}

\title{Timed Orchestration of Component-based Systems}

\author{Chih-Hong Cheng\inst{1}\thanks{Part of this work has been initiated at ABB Research.}%
	\and L\u acr\u amioara~A\c stef\u anoaei\inst{1} \and Harald Ruess\inst{1}
	\and \\
	Souha Ben-Rayana\inst{2} \and Saddek Bensalem\inst{2}}

\institute{
	fortiss - An-Institut Technische Universit{\"a}t  M\"{u}nchen\\
	\and
	Verimag Laboratory, Grenoble, France
}

\begin{document}

\maketitle

 \begin{abstract}
 	
 	Individual machines in flexible production lines explicitly
        expose capabilities at their interfaces by means of parametric
        {\em skills} (e.g. drilling)\@. Given such a set of configurable machines,
        a line integrator is faced with the problem of finding and
        tuning parameters for each machine such that the overall
        production line implements given safety and temporal
        requirements in an optimized and robust fashion.  We formalize
        this problem of configuring and orchestrating flexible
        production lines as a parameter synthesis problem for systems
        of {\em parametric timed automata}, where interactions are
        based on skills. Parameter synthesis problems for
        interaction-level LTL properties are translated to parameter
        synthesis problems for state-based safety properties.  For
        safety properties, synthesis problems are solved by checking
        satisfiability of $\exists\forall$SMT constraints. For
        constraint generation, we provide a set of computationally
        cheap over-approximations of the set of reachable states,
        together with fence constructions as sufficient conditions for
        safety formulas. We demonstrate the feasibility of our
        approach by solving typical machine configuration problems as
        encountered in industrial automation.
 \end{abstract}

\section{Introduction}
We consider the problem of automatically configuring and orchestrating a 
set of production machines with standardized interfaces.
For example, machine interfaces in the packaging industry are expressed in the standardized PackML\footnote{\url{http://www.omac.org/content/packml}} notation,
and {\em skill sets} such as \sig{fill-box} or \sig{drill} have recently been 
introduced, in the context of flexible production lines of the Industrie 4.0 programme, for describing parametric machine 
capabilities~\cite{capability}\@.\footnote{\url{http://www.autonomik40.de/en/OPAK.php}}

Given such a set of configurable machines, a production line integrator is faced with the task of finding and tuning parameters for each machine such that the overall production line 
satisfies required safety and temporal constraints.
Typical line requirements from the practice of industrial automation include, for example, line-level safety, 
error-handling, and the orchestrated execution of sequences of skills intermixed with machine-to-machine communication 
primitives. In addition, production lines are usually required to perform in an optimized and robust manner.


We tackle this problem of orchestrating and configuring parametric
production systems by means of parameter synthesis problems for
systems of interacting {\em parametric timed automata} (PTAs), where
multi-party interactions between individual PTAs represent skills and
machine-to-machine communication.

In a first step, parameter synthesis problems for interaction-level
linear temporal logic (LTL)~\cite{pnueli1977temporal} properties are
translated, based on constructions in bounded
synthesis~\cite{ScheweF07a,acacia12,Ehlers11}\@, into parameter
synthesis problems for state-based safety properties. The key element
here is the construction of a deterministic monitor similar to bounded
LTL synthesis.  Due to the use of clocks, however, there are some
technical differences to this well-known construction, including a
different upper bound of the maximum number of required unrolling
steps.  Whenever parameters are integer bounded, we demonstrate the
existence of a sufficient upper bound for unrolling the negated
property automata, such that one can conclude that no parameter
assignment can guarantee the specified LTL property.

  Then, parameter synthesis problems for safety properties are transformed
  to solving $\exists\forall$SMT satisfiability problems of the form
  $\exists x :\, \forall y: \Reach(x, y) \rightarrow (\neg
  \phi_{deadlock}(x,y)\wedge \rho_{\mathit{safe}}(x,y))$, where $x$
  represents the set of parameters to be synthesized, $y$ represents
  all the component states including local clocks, $\Reach$ represents the set of
  reachable states, $\neg \phi_{deadlock}(x,y)$ denotes deadlock freeness, 
  and $\rho_{\mathit{safe}}$ denotes the required safety condition.  
  In general, the computation
  of the parametric image $\Reach$ is undecidable for parameters of
  unbounded domain~\cite{alur93}\@.  For bounded (integer) parameters,
  however, $\Reach$ can be computed precisely by enumerating all
  valuations of parameters and, subsequently, constructing the region
  graph for each parameter valuation\@.  Usually, zone or region
  diagrams~\cite{jovanovic13:synthesis-pta,henzinger94} are holistic (computationally expensive) approaches used to
  compute precise images for parameters of bounded domain or abstraction for parameters of unbounded domains.  Instead, we are proposing
  a set of computationally-cheap over-approximations of $\Reach$ for avoiding eager and expensive computations of  $\Reach$\@. Novel constructions include over-approximations
  based on finite depth interaction-history and fence constructions
  for guaranteeing safety. We also demonstrate the usefulness of these over-approximations with examples based on 
  flexible production systems.
 
 Due to the proposed reduction of parametric synthesis problems to general
 $\exists\forall$SMT formulas, one may encode and simultaneously solve both qualitative and
 quantitative (e.g., \sig{min}, lexicographic) requirements on synthesized solutions.
Moreover, the $\exists\forall$-centric encoding of this paper also allows for the
 synthesis of non-timing parameters.  Our use of two SMT solvers for
 solving $\exists\forall$SMT is an extension of using two SAT solvers
 for solving 2QBF formula~\cite{2SAT}\@. The new approach here is to
 exploit this decoupling to also integrate quantitative aspects in solving synthesis problems.
 
To validate our approach, we have implemented a prototype which
includes an $\exists\forall$ constraint generator 
and an $\exists\forall$ constraint solver (\sig{EFSMT}).  Our initial
experiments are encouraging in that our prototype implementation
reasonably deals with synthesis problems from our benchmark set with
$20$ unknown parameters and $10$ clocks; that is, the proposed synthesis
algorithms seems to be ready to handle the fully automatic orchestration
of, at least, smaller-scale modular automation systems.

\vspace{1mm}
\noindent\textbf{Related Work.}  
Verification and synthesis of parametric timed automata have recently
been considered, among others,
by~\cite{hune02:mc-pta,andreS11:synt-pta,jovanovic13:synthesis-pta}\@.
These techniques have also been implemented in the tools
IMITATOR~\cite{imitator} and Romeo~\cite{romeo}, which search for
constraints on parameters for guaranteeing the existence of a
bisimulation between any timed automata (TA) satisfying the constraints and an initial instantiation of the input PTA\@. 
One of the main differences between solving strategies centers around
forward versus backwards search, as Romeo starts, using a CEGAR-like
strategy, from a counterexample, whereas IMITATOR starts from a {\em
good} initial valuation of the parameters. In contrast, we are
  finding the right parameter values which guarantee that the system
  is deadlock free, and satisfies state-based and interaction-level
  properties. 
  Existing approaches, which are based on computing and exhaustively exploring the global state space, usually do not perform
  well even for relatively simple properties such as deadlock-checking, and their implementations are currently restricted to handle 
  problems with only a relatively small number  (in the order of ten)  automata.
  contrast, we apply a constraint-based solving approach and use a
  number of compositional techniques for generating local timing
  invariants for efficiently solving $\exists\forall$-formulae with
  \textsf{EFSMT}\@.  Apart from scalability, the
  $\exists\forall$-centric approach also allows for the integration of
  quantitative objectives.  Finally, to the best of our knowledge,
  current verification and synthesis tools such
  UPPAAL~\cite{bdl:2004:uppaal,behrmann2007uppaaltiga}, IMITATOR, or
  Romeo do not support neither multi-party interactions nor
  qualitative interaction-level properties (LTL)\@.

  
%



\vspace{1mm}

\noindent\textbf{Organization of the paper.} 
Section~\ref{sec.background} recalls the basic definitions for PTAs, 
safety and transaction-level properties for interacting systems of PTAs, and the orchestration problem for these systems of PTAs\@.
The main technical developments for solving timed orchestration synthesis are presented in Section~\ref{sec.workflow}\@. 
 Section~\ref{sec.evaluation} provides some experimental results with a prototype implementation.  Final conclusions are 
summerized in Section~\ref{sec.summary}\@. 



\section{Parametric Component-based Systems and Properties}\label{sec.background}

\begin{figure}[t]
	\centering
	\includegraphics[width=\columnwidth]{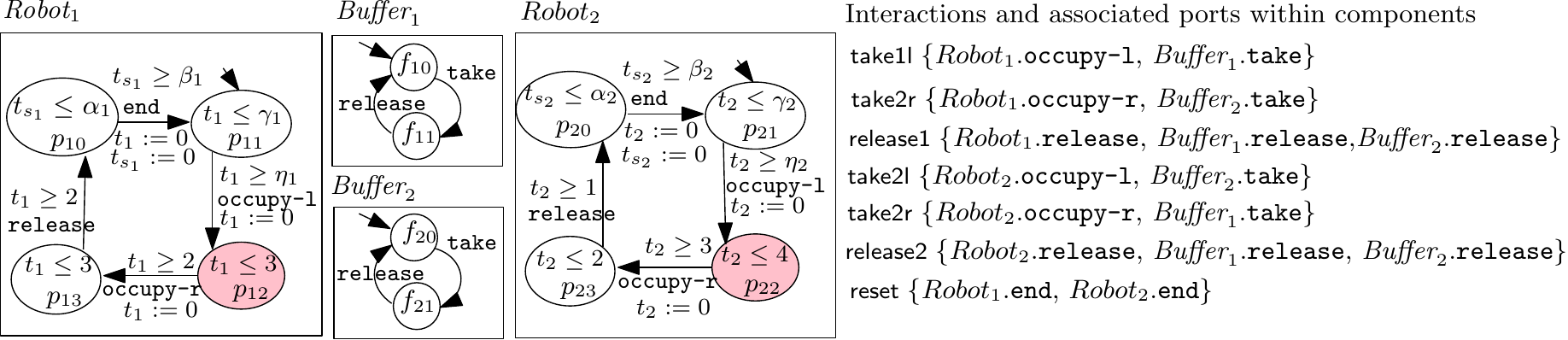}

	\caption{Parametric robot stations with shared buffer locations.}

	\label{fig:TimedPhilosopher}
\end{figure}

We briefly review some basic notions for systems of parametric timed automata, and formally state the problem of timed orchestration synthesis.
\begin{definition}[Component]
A component $C (Q, q, X, P, $ $Jump, Inv)$ is a parametric timed
automaton, where:
\begin{itemize}
	\item $Q$ is a finite set of \emph{locations}, and $q \in Q$ is the \emph{initial location}
	\item $X$ is the set of	\emph{clock variables}
	\item $P$ is a finite set alphabet called ports (edge labels)
	\item $\rightarrow \subseteq Q\times Guards \times P
          \rightarrow Q \times Resets$ is the set of \emph{discrete
            jumps} between locations. $Guards$ is the conjunction of
          inequalities of the form $x \sim k$; $Resets \subseteq X$ is
          a set of clock variables to be reset after discrete jump. We
          assume that every port $p\in P$ is associated with only one
          discrete jump in $Jump$
	\item $Inv$ is the set of \emph{location conditions} mapping
          locations to conjunctions of disequalities of form $x \leq
          k$
\end{itemize}
with $x \in X$, $k\in \mathbb{N}_0 \cup V$ and $\sim\; \in\{=,>,\geq\}$. 
\end{definition}
For ease of reference, we use the notation $C.p$ to denote the
port $p$ of component $C$, as shown in
Fig.~\ref{fig:TimedPhilosopher}\@.
\begin{definition}[System] A \emph{system} is a tuple $\mathcal{S} = (V, C, \Sigma, \Delta)$,  where:
\begin{itemize}
  \item $V$ is a finite set of \emph{unknown parameters}
  \item $C = \bigcup_{i=1}^m C_i$ is a finite set of \emph{components}
  \item $\Sigma$ is a finite set of system-level events
    (\emph{interactions}), called \emph{interaction alphabet}.
  \item $\Delta: \Sigma \rightarrow \bigcup_{i=1}^m P_i$ associates
    each interaction $\sigma$ with some ports within components.
We assume that every port is associated with at least one interaction.
\end{itemize}
\end{definition}        

\comments{
\begin{table}[t]

	\centering
	\begin{tabular}{|l|l|}
		\hline
		\sig{take1}	& {\scriptsize \{$\phio_1.\texttt{takeleft}$, $Fork_1.\texttt{take}$\}} \\
		\sig{eat1}	& {\scriptsize \{$\phio_1.\texttt{eat}$, $Fork_2.\texttt{take}$\}} \\
		\sig{release1}	& {\scriptsize \{$\phio_1.\texttt{release}$, $Fork_1.\texttt{release}$, $Fork_2.\texttt{release}$\}} \\
		\sig{take2}	& {\scriptsize\{$\phio_2.\texttt{takeleft}$, $Fork_2.\texttt{take}$\}} \\
		\sig{eat2}	& {\scriptsize\{$phio_2.\texttt{eat}$, $Fork_1.\texttt{take}$\}}  \\
		\sig{release2}	& {\scriptsize\{$\phio_2.\texttt{release}$, $Fork_1.\texttt{release}$, $Fork_2.\texttt{release}$\}} \\
		\sig{reset}	& {\scriptsize\{$\phio_2.\texttt{end}$, $\phio_2.\texttt{end}$\}}\\
		\hline
	\end{tabular}
	\vspace{2mm}
	\caption{$\Delta$ associating interactions with ports for Figure~\ref{fig:TimedPhilosopher}.}
	\label{delta}
\end{table}
}

The concrete semantics of a system under a valuation of the unknown
parameters follows the standard semantics of timed
automata~\cite{alur1994tta}, except that discrete jumps are
synchronized by interactions (see \cite{AstefanoaeiRBBC14} for
details). A time run is a maximal sequence of transitions $(\vec{q}_0,
\vec{v}_0) \stackrel{\sigma_0}{\rightarrow} (\vec{q}_1,
\vec{v}_1) \stackrel{\sigma_1}{\rightarrow} \dots (\vec{q}_n,
\vec{v}_n) \stackrel{\sigma_n}{\rightarrow} (\vec{q}_{n+1}, \vec{v}_{n+1}) \dots$ where
$\vec{q}_i$ denotes a location in the system $\mathcal{S}$, $\sigma_i$ is an
interaction and $\vec{v}_i$ is a valuation of the clocks in $\mathcal{S}$.

For the ease of reference, we introduce the following notations.  For
$\sigma \in \Sigma$, we denote $\tEn{\sigma}$ to be the necessary
condition for enabling a location combination to trigger $\sigma$ by
only allowing \emph{finite-time} evolving, where the definition of $\tEn{\sigma}$ is taken from~\cite{tripakis99:progress}. 
If from a location $q$ one can delay the triggering of $\sigma$ indefinitely, then $\tEn{\sigma}$ for that location is by default \sig{false}.
 Given a valuation $\vec{v}$ assigning the variables in $V$,
 $\mathcal{S}(\vec{v})$ denotes the resulting concrete timed system
 and $\tEn{\sigma}(\vec{v})$ denotes the resulting constraint of
 enabling conditions. 
 For infinite time runs $\rho$ with infinite discrete jumps,
  we use $\rho_{\Sigma}$ to denote the corresponding
 $\omega$-word with symbols from the interaction alphabet.

Figure~\ref{fig:TimedPhilosopher} illustrates these concepts by means of  a variation of the resource contention problem in terms of timed-based control over robots,
which is used as a running example.  
\begin{example}
\label{eg:philo1}
Given $n$ robots, robot~$i$ first
accesses buffer~$i$ then buffer $(i-1)\% n$. Figure~\ref{fig:TimedPhilosopher} depicts the system for $n=2$,
with the set of unknown parameters
$V=\{\alpha_1, \alpha_2,\beta_1,$ $\beta_2, \gamma_1, \gamma_2, \eta_1, \eta_2\}$\@.
 Each $\phio_i$ has four
ports \{\port{occupy-l}, \port{occupy-r}, \port{release}, \port{end}\}.
This system has the interactions
$\Sigma=\{\sig{take1l}, \sig{take1r}, \sig{release1}, \sig{take2l},$
$\sig{take2r}, \sig{release2},\sig{reset}\}$, and $\Delta$ is defined to the
right of Figure~\ref{fig:TimedPhilosopher}\@. For
$\tEn{\sig{release1}}$, the necessary condition for
interaction \sig{release1} to eventually take place without discrete
jumps, is $p_{13} \wedge f_{11} \wedge f_{21} \wedge (t_1\leq 3) \wedge
(2 - t_1 \leq 3 - t_1) \wedge (3 - t_1 \leq \alpha_1 - t_{s_1})
\wedge \stay{\phio_2}$. The trivial condition $2 - t_1 \leq 3 - t_1$ is to guarantee that the minimum required time for $t_1$ to have the guard enabled does not let the location invariant of $p_{13}$ be violated. Constraint  $3 - t_1 \leq \alpha_1 - t_{s_1}$
is to ensure that the latest delay for enabling the transition, i.e., time elapse of $t_1$ to reach the boundary of invariant (which is larger than $2 - t_1$, the shortest delay required to enable the guard), is less
than the time it takes $t_{s_1}$ to reach $\alpha_1$.
This makes it possible to
jump to location~$p_{10}$. Constraint $\stay{\phio_2} = (p_{20} \wedge
3 - t_1 \leq \alpha_2 - t_{s_2}) \vee (p_{21} \wedge 3 -
t_1 \leq \gamma_2 - t_2) \vee (p_{22} \wedge 3 - t_1 \leq 4 - t_2)
\vee (p_{23} \wedge 3 - t_1 \leq 2 - t_2)$ is to ensure
that $\phio_2$ is able to stay within its location, before the discrete jump is taken. Note that the
clock condition at $p_{13}$ involved in the interaction \sig{release1}
ensures that time cannot be delayed at infinity.

Now, consider the assignment 
$\vec{v}\stackrel{\triangle}{=} \{\alpha_1=\beta_1=\alpha_2=\beta_2 = 30, \gamma_1=5, \gamma_2=20,
\eta_1=0, \eta_2=15\}$\@, which results in an infinite behavior on the interaction level, as presented
by the $\omega$-word $\rho_{\Sigma}$: $(\sig{take1l}\, \sig{1take1r}\,
\sig{release1}\, \sig{take2l}, \sig{take2r}\, \sig{release2}\,
\sig{reset})^{\omega}$\@.
\end{example}

\begin{definition}[Properties] 
\label{def:props}
We consider three types of properties:
\begin{itemize}
	\item \emph{Component-level properties} $\phi_C$ are constraints over $V$.


	\item \emph{Safety properties} $\phi_{state}$ are state properties to be satisfied in every reachable state of the system.  Typically, they are location-wise and express relations between clocks. 
	
	\item \emph{Interaction-level properties} $\phi_{int}$ are LTL
          specifications over $\Sigma$. A concrete timed system
          $\mathcal{S}(\vec{v})$ satisfies $\phi_{int}$ iff every time
          run $\rho$ of $\mathcal{S}(\vec{v})$ involves infinitely
          many discrete jumps, and the corresponding $\omega$-word
          $\rho_{\Sigma}$ is contained in $\phi_{int}$  by standard LTL semantics.
	
	
	
\end{itemize}
\end{definition}

\begin{example}
Consider the following properties to be
synthesized for the robot running example as displayed in Figure~\ref{fig:TimedPhilosopher}\@:
\begin{itemize}
	\item All parameters should be within $[0, 30]$
          [Component-level property].
	\item Deadlock freedom [Safety property]. 
	\item $t_1+t_2$ should always be less than 60
           [Safety property]. 
	\item Promptness / exclusiveness: $\phi_{prompt}:=
          \bigwedge_{i} \G(\sig{take}_i\sig{l} \rightarrow \textbf{X}
          \bigwedge_{j\neq i}\neg\sig{take}_j\sig{l})$, i.e., disallow
          $\phio_j$ to perform $\sig{take}_j\sig{l}$ immediately after
          $\sig{take}_i\sig{l}$ from $\phio_i$ [Interaction-level property].
\end{itemize}
\end{example}
\begin{definition}[Timed Orchestration Synthesis] 
\label{def:tos}
 Given $\mathcal{S}=(V, C,\Sigma, \Delta)$ and properties $\phi_C$, $\phi_{state}$,
$\phi_{int}$, the problem of \emph{timed orchestration synthesis} is to find an assignment
$\vec{v}$ for $V$ such that $\vec{v}$ satisfies $\phi_C$, and $\mathcal{S}(\vec{v})$ satisfies 
both $\phi_{state}$ and $\phi_{int}$; such a satisfying assignment $\vec{v}$ is also called a \emph{solution}. 
\end{definition}
For example, the assignment given in Example~\ref{eg:philo1} is a solution for timed orchestration
synthesis when applied to our running example.





\section{Timed Orchestration Synthesis}\label{sec.workflow}

This section describes our main constructions for solving timed orchestration synthesis problems.
We first translate timed orchestration synthesis problems for LTL
properties to corresponding synthesis problems for safety properties
(Sec.~\ref{sub.sec.monitor})\@. Second,
$\exists\forall$SMT constraints are generated for the latter problem, whereby existential 
variables quantify over the parameters to be synthesized and universal variables quantify over system states
(Sec.~\ref{sub.sec.parameterized.invariants}). 
Third, the 
$\exists\forall$SMT constraints are solved by means of two alternating quantifier-free SMT solvers (Sec.~\ref{sub.sec.efsmt}) for each polarity\@.
In order to simplify the exposition below, we omit $\phi_C$ as it ranges only over the existentially-quantified parameters  in $V$, and 
concentrate on the properties $\phi_{state}$ and $\phi_{int}$\@.

\subsection{Transforming Interaction-level to Safety Properties}\label{sub.sec.monitor}

To effectively synthesize parameters such that interaction-level
properties $\phi_{int}$ are satisfied, we adapt bounded
LTL synthesis~\cite{ScheweF07a} to our context. The underlying
strategy is to construct a deterministic progress monitor from
$\phi_{int}$.  The monitor is meant to keep track of the final states
visited in the B\"uchi automaton $\mathcal{A}_{\neg \phi_{int}}$
corresponding to $\neg\phi_{int}$ during system execution. To achieve
this, we equip the monitor with a dedicated \sig{risk} state
representing that a final state in $\mathcal{A}_{\neg \phi_{int}}$ has
been visited for $k$ times. When the risk state is never reached for
all possible runs, all final states in $\mathcal{A}_{\neg \phi_{int}}$
are visited finitely often (i.e., less than $k$ times). This
observation is sufficient to conclude that the system satisfies
$\phi_{int}$. This is the intuition behind Algorithm~\ref{alg:int}.

Algorithm~\ref{alg:int} uses $\Sigma_{\phi_{int}} \subseteq \Sigma$ to
be the set of interactions from $\phi_{int}$ and $\#$ as a symbol
not within $\Sigma$. 
On Line~\ref{l2}, the symbol $\#$ is used to mark labels corresponding
to interactions $\sigma$ not appearing in $\pintt$. On Line~\ref{l3} a
deterministic progress monitor is constructed $\mon{k}$ by unrolling
$\mathcal{A}_{\neg\phi_{int}}$ via function $\textbf{monitor}(\mathcal{A}_{\neg\phi_{int}}, k)$, which is similar to the
approach in bounded LTL synthesis~\cite{ScheweF07a}. Consequently, we
omit it and instead provide a high-level description of what it does (see below example for understanding):
Starting from the initial state of $\mathcal{A}_{\neg\phi_{int}}$,
$\Sigma_{\phi_{int}}\cup \{\#\}$ is used to \emph{unroll} all traces
of $\mathcal{A}_{\neg\phi_{int}}$ and to create a deterministic
monitor $C_{\neg \phi_{int},k}$. Each location\footnote{To avoid
  ambiguity, we call a state in the monitor component ``location''
  while keeping the name ``state'' for B\"uchi automaton.}  in
$C_{\neg \phi_{int},k}$ records the set of states being visited in the
B\"uchi automaton. For each location, the number of times a final state
in $\mathcal{A}_{\neg\phi_{int}}$ has been visited previously is counted.
The algorithm maintains a queue of unprocessed locations. For each
unprocessed location in the queue, every interaction $\sigma \in
\Sigma_{\phi_{int}} \cup \{\#\}$ is selected to create a successor
location respectively. A state $s'$ is stored in the successor
location, if state $s$ is in the unprocessed location and if in the
post-processed B\"uchi automaton, a transition from $s$ to $s'$ via
edge labeled $\sigma$ exists. In addition, the number of visited final
states is updated. Whenever a final state in
$\mathcal{A}_{\neg\phi_{int}}$ has been visited $k$ times, the unroll
process replaces the location of $C_{\neg \phi_{int},k}$ by
\sig{risk}, a dedicated location with no outgoing edges. 

\begin{figure}[htp]
\centering
\includegraphics[width=\columnwidth]{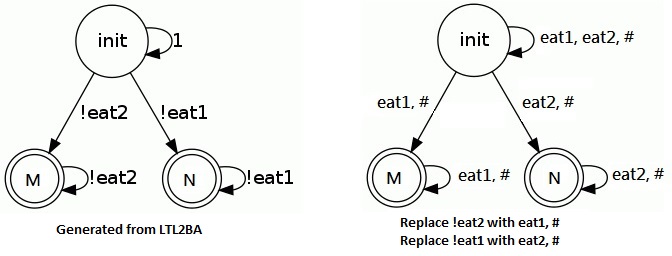}
\caption{(a)(b) B\"uchi automaton for $\neg \phi_{prompt}$ before and
  after postprocessing. (c) 
$\mathcal{C}_{\neg \phi_{prompt}, 1}$.
}
\label{fig:negated.property}
\end{figure}

Once the monitor is constructed, an augmented system $\mathcal{S}_{inv,k}$ is
created from $\mathcal{S}$ (Line~\ref{l6}). The interaction set in the
augmented system $\mathcal{S}_{inv,k}$ is the one from Line~\ref{l5}
where all property-unrelated interactions $\sigma$ are marked with
$\#$. Finally, on Line~\ref{l7} the state predicate $\phi_{deadlock}$
expressing the deadlock condition is constructed from the new set of
interactions.
\vspace*{.3cm}
\begin{algorithm}
\vspace*{.1cm}
\caption{Translate $\pintt$ into $\deadd$ and construct a monitored system $\systik$}\label{alg:int}
\vspace*{.1cm}
\vspace*{.1cm}
\begin{algorithmic}[1]
\Statex \textbf{Input: }$\syst$, $\pintt$, $k$
\Statex \textbf{Output: }$\systik, \deadd$
\vspace*{.1cm}
\State construct a B\"uchi automaton $\mathcal{A}_{\neg\phi_{int}}$ for the negated property of  $\phi_{int}$\label{l1}
\State postprocess $\mathcal{A}_{\neg\phi_{int}}$ by replacing every label $\neg \sigma$ with $\Sigma_{\phi_{int}}\setminus \{\sigma\} \cup \{\#\}$\label{l2}
\State $\mon{k} := $ \textbf{monitor}($\mathcal{A}_{\neg\phi_{int}}, k$) 
\label{l3}
\State $\Delta_{inv,k}(\sigma):=$ {$\sigma \in \Sigma_{\phi_{int}}$} ? {$\Delta(\sigma)\cup\{C_{\neg\phi_{int},k}.\sigma\}$} : {$\Delta(\sigma)\cup\{C_{\neg\phi_{int},k}.\#\}$} \label{l5}
\State $\mathcal{S}_{inv,k} = (V, C\cup \{C_{\neg\phi_{int},k}\}, \Sigma, \Delta_{inv,k})$ \label{l6}
\State $\phi_{deadlock} :=\bigwedge_{\sigma  \in \Sigma} \neg \tEn{\sigma}$  \label{l7}
\State \textbf{return } $\mathcal{S}_{inv,k}$, \text{ } $\deadd$
\end{algorithmic}
\end{algorithm}
\mycomment{
\vspace*{.3cm}
\begin{algorithm}
\vspace*{.1cm}
\caption{Deterministic Progress Monitor}\label{alg:mon}
\vspace*{.1cm}
\vspace*{.1cm}
\begin{algorithmic}[1]
\Statex \textbf{Input: }$\mathcal{A}_{\neg\phi_{int}}$, $k$
\Statex \textbf{Output: }$\mon{k}$
\vspace*{.1cm}
\State todo: extract an algo from the description above
\State $tovisit.push(q_0)$ 
\State ...
\end{algorithmic}
\vspace*{.3cm}
\end{algorithm}
\vspace*{.3cm}
}
\begin{example}
We illustrate the steps of the algorithm~\ref{alg:int} using the robot running example. 
Figure~\ref{fig:negated.property}-(a) illustrates the
result $\mathcal{A}_{\neg \phi_{prompt}}$ for property $\phi_{prompt}$
(Line~\ref{l1}), and  (b) displays the result after post-processing
(Line~\ref{l2})\@. 

To illustrate the result of unrolling in Line~\ref{l3},
Figure~\ref{fig:negated.property}-(c) shows it for $k=1$. There, the
initial location stores \{$s_0$[$s_3$(0)]\}, where [$s_3$(0)] is to
indicate that at $s_0$, one has not yet reached $s_3$ previously. When
the initial location \{$s_0$[$s_3(0)$]\} takes interaction
\sig{take1l}, it goes to \{$s_0$[$s_3(0)$], $s_1$[$s_3(0)$]\}, as in
Figure~\ref{fig:negated.property}-(b), state $s_0$ can move to $s_0$
or $s_1$. Notice  that it a destination location has possibly been created previously. 
For example, in
Figure~\ref{fig:negated.property}-(c), for the initial location
\{$s_0$[$s_3(0)$]\} to take interaction \#, it goes back to
\{$s_0$[$s_3(0)$]\}. 
For
\{$s_0$[$s_3(0)$], $s_1$[$s_3(0)$]\} to take interaction \sig{take2l},
it moves to a new location \{$s_0$[$s_3(0)$], $s_3$[$s_3(1)$],
$s_2$[$s_3(0)$]\}. This new location is then replaced by \sig{risk},
as in this example, we have $k=1$.

As for the new interaction set in the monitored system, we show two
examples with respect to whether the interaction is in $\phi_{prompt}$:
\begin{align*}
& \Delta_{inv,k}(\sig{take1l}) = \{\phio_1.\port{occupy1-l}, \buffer_1.\port{take}, C_{\neg\phi_{prompt}}.\port{take1l}\}\\
& \Delta_{inv,k}(\sig{take2r}) = \{\phio_2.\port{occupy2-r}, \buffer_1.\port{take}, C_{\neg \phi_{prompt}}.\#\}
\end{align*}

\end{example}	

Notice that the introduction of $\#$ symbol simplifies the unroll construction in bounded synthesis. 
Another difference to vanilla bounded synthesis is that, in the context of unrolling, every state
has outgoing edges of size $|\Sigma_{\phi_{int}}\cup \{\#\}|$. 
In contrast, in  bounded LTL synthesis, each $\sigma$ is viewed as a Boolean
variable, which creates, in the worst case, on the order of $2^{|\Sigma|}$ outgoing edges.

The following result reduces timed orchestration synthesis for interaction-level properties to a corresponding timed orchestration synthesis problem on state-based properties only.
\begin{lemma}
\label{lem:correct}
  Given $\vec{v}$ an assignment of $V$, $\mathcal{S}(\vec{v})$ satisfies $\phi_{int}$ if all time runs of $\mathcal{S}_{inv,k}(\vec{v})$ reach neither the location \sig{risk} in $C_{\neg\phi_{int},k}$ nor a state where $\phi_{deadlock}(\vec{v})$ holds. 

\end{lemma}

\proof (Sketch) Assume that any time run $\rho$ in
$\mathcal{S}_{inv,k}(\vec{v})$ does not visit location \sig{risk} or
any state where $\phi_{deadlock}(\vec{v})$ holds. We need to show that
$\mathcal{S}(\vec{v})$ satisfies $\phi_{int}$.
\begin{enumerate}
\item 	Because $\dead$ does not hold and because time runs are maximal, any such time run $\rho$ is infinite. 
%
\item From an infinite time run $\rho$, we show that $\rho$ defines an
  $\omega$-word $\rho_{\Sigma}$: as $\phi_{deadlock}(\vec{v})$ is
  never reached, $\bigvee_{\sigma \in \Sigma} \tEn{\sigma}(\vec{v})$
  is an invariant for all reachable states. Recall that $\tEn{\sigma}$
  is the necessary condition for enabling a location to trigger
  $\sigma$ by only allowing \emph{finite-time} evolving. Therefore,
  for all reachable states, one of the interaction (discrete jump)
  must appear after finite time. Thus, $\rho$ contains infinitely many
  discrete jumps and consequently, $\rho$ defines an $\omega$-word
  $\rho_{\Sigma}$.
\item By construction, every location in the monitor has edges labeled
  in $\sigma \in \Sigma_{\phi_{int}} \cup \{\#\}$, and \# marks each
  property-unrelated interaction in $\Sigma\setminus
  \Sigma_{\phi_{int}}$ (Line~\ref{l5}). From this observation,
  together with the fact that $\mathcal{S}_{inv,k}(\vec{v})$ does not
  restrict the behavior of $\mathcal{S}(\vec{v})$, we have that a time
  run $\rho$ in $\mathcal{S}_{inv,k}(\vec{v})$ not reaching \sig{risk}
  is bisimilar to a time run $\rho'$ in $\mathcal{S}(\vec{v})$, with
  $\rho$ and $\rho'$ defining the same $\omega$-word $\rho_{\Sigma}$.
\item Recall that $C_{\neg\phi_{int},k}$ is an unroll of
  $\mathcal{A}_{\neg\phi_{int}}$. From this, together with the
  existence of $\rho_{\Sigma}$ and the fact that while running
  $\rho_{\Sigma}$ in $\mathcal{A}_{\neg\phi_{int}}$ no final state is
  reached infinitely many times, we have that $\rho_{\Sigma}$ does not
  satisfy $\neg\phi_{int}$. Consequently, we can conclude that for
  every time run $\rho'$ in $\mathcal{S}(\vec{v})$, the corresponding
  $\rho_{\Sigma}$ satisfies $\phi_{int}$. \qed
\end{enumerate}
\newcommand{\mk}{k^*}
By Lemma~\ref{lem:correct}, it is sufficient to only consider safety
properties when performing orchestration synthesis.
Notice howerver that,  if for a given fixed $k$ Lemma~\ref{lem:correct} fails for all possible
assignments $\vec{v}$ then one may not conclude that no solution exists for
the orchestration synthesis problem as there might be a larger $k$ for
which Lemma~\ref{lem:correct} does hold.


Next we show that it is futile to go beyond a reasonable bound. More precisely, if the domains of the parameters in the input system $\mathcal{S}$ are bounded, then one can effectively compute a limit $\mk$ on $k$ such that: if Lemma~\ref{lem:correct} is not applicable for $\mk$ then it is also not applicable for any strictly larger $k$\@. 

\begin{lemma}
\label{lem:k}
Let all parameters in $\syst$ have bounded integer domains with common
upper bound $\lambda$ and $\delta$ be the number of regions in $\syst$
when all parameters within the location and guard conditions are
assigned $\lambda$. Let $|\mathcal{A}_{\neg\phi_{int}}|$ be the number
of locations in $\mathcal{A}_{\neg\phi_{int}}$, and $\eta$ be the
number of discrete location combinations in $\mathcal{S}$, i.e., $\eta
= |Q_1||Q_2|\dots|Q_m|$. Finally, let $\mk =
\delta^2\eta^2|\mathcal{A}_{\neg\phi_{int}}||\Sigma|+1$.

Given $\vec{v}$ an
assignment of $V$, if there exists a time run of
$\mathcal{S}_{inv,\mk}(\vec{v})$ reaching either \sig{risk} in
$C_{\neg\phi_{int},\mk}$ or a state where $\phi_{deadlock}(\vec{v})$
holds, then $\mathcal{S}(\vec{v})$ does not satisfy $\phi_{int}$.

		


\end{lemma}

\proof (Sketch) Let $\rho$ be a time run of
$\mathcal{S}_{inv,\mk}(\vec{v})$ which reaches either \sig{risk} in
$C_{\neg\phi_{int},\mk}$ or a state where $\phi_{deadlock}(\vec{v})$
holds. We have two cases:
\begin{itemize}
\item \textbf{Case 1}: \sig{risk} is not reached, equally, $\rho$ reaches a state where $\phi_{deadlock}(\vec{v})$ holds. The deadlock of $\mathcal{S}_{inv,\mk}(\vec{v})$ under $\rho$ is irrelevant to the monitor component, as the monitor component does not hinder any execution apart from \sig{risk}.  Therefore, $\mathcal{S}(\vec{v})$ contains deadlock states, and 
 $\mathcal{S}(\vec{v})$ does not satisfy $\phi_{int}$, as reaching a deadlock state means that one can not create an  $\omega$-word from that time run. 
	
	
	
	
	\item \textbf{Case 2}: \sig{risk} is reached in the $\mk$-th
          unroll. We need to show that with a prefix of a violating
          time run $\rho$ in $\mathcal{S}_{inv,\mk}(\vec{v})$ that
          visited one final state $s$ in
          $\mathcal{A}_{\neg\phi_{int}}$ for $\mk$ times, one proves,
          by tailoring a fragment of $\rho$, the existence of a time
          run $\rho'$ in $\mathcal{S}(\vec{v})$ such that the
          $\omega$-word of $\rho'$, when applying to
          $\mathcal{A}_{\neg\phi_{int}}$, guarantees to visit $s$
          arbitrary many times.  This is done with the help of two results:
          (1) the number of regions in a timed automaton is finite,
          and (2) the pigeonhole principle.
          concrete values, the system is a timed automaton and the
          number of regions is finite. The total number of regions is
          bounded by $\delta$. Recall that the executions in $C_{\neg
            \phi_{int},k}$ reflect executions in the B\"uchi automaton
          $\mathcal{A}_{\neg\phi_{int}}$. Consider a discrete jump in
          the system. With a specific destination state $s$ in
          $\mathcal{A}_{\neg\phi_{int}}$, one can actually capture a
          discrete jump in $\mathcal{S}_{inv,k}(\vec{v})$ by only
          viewing its change in regions and locations. To reflect such
          changes, we use tuples $\langle(r_{source}, l_{source},
          s_{source}),\sigma,(r_{dest}, l_{dest}, s)\rangle$, where
          $r_{source}$ and $r_{dest}$ are source and destination
          regions in $\mathcal{S}(\vec{v})$, $l_{source}$ and
          $l_{dest}$ are source and destination location in
          $\mathcal{S}(\vec{v})$, $s_{source}$ is the source state in
          $\mathcal{A}_{\neg\phi_{int}}$ with interaction $\sigma \in
          \Sigma$.  The total number for all such $\langle(r_{source},
          l_{source}, s_{source}),\sigma,(r_{dest}, l_{dest},
          s)\rangle$ tuples is bounded by $\delta^2 \eta^2
          |\mathcal{A}_{\neg\phi_{int}}| |\Sigma|$.  Therefore, when
          the violating $\rho$ visits a particular final state $s$ in
          $\mathcal{A}_{\neg\phi_{int}}$ for $\mk$ times, in the
          corresponding region representation, one particular tuple
          $\langle(r_{source}, l_{source},
          s_{source}),\sigma,(r_{dest}, l_{dest}, s)\rangle$ should
          have appeared twice (due to pigeonhole principle). The clock
          valuations associated to $l_{source}$, resp. $l_{dest}$ may
          be different in the two tuples. However, the corresponding
          states are region-equivalent and consequently
          bisimilar. Thanks to this, $\mathcal{S}$ can evolve
          region-bisimilarly until the tuple $\langle(r_{source},
          l_{source}, s_{source}),\sigma,(r_{dest}, l_{dest},
          s)\rangle$ appears for the third time, and so on. While
          repeating this pattern a time run visiting $s$ infinitely
          often is constructed.

\end{itemize}



\begin{figure}
	\centering
	\includegraphics[width=\columnwidth]{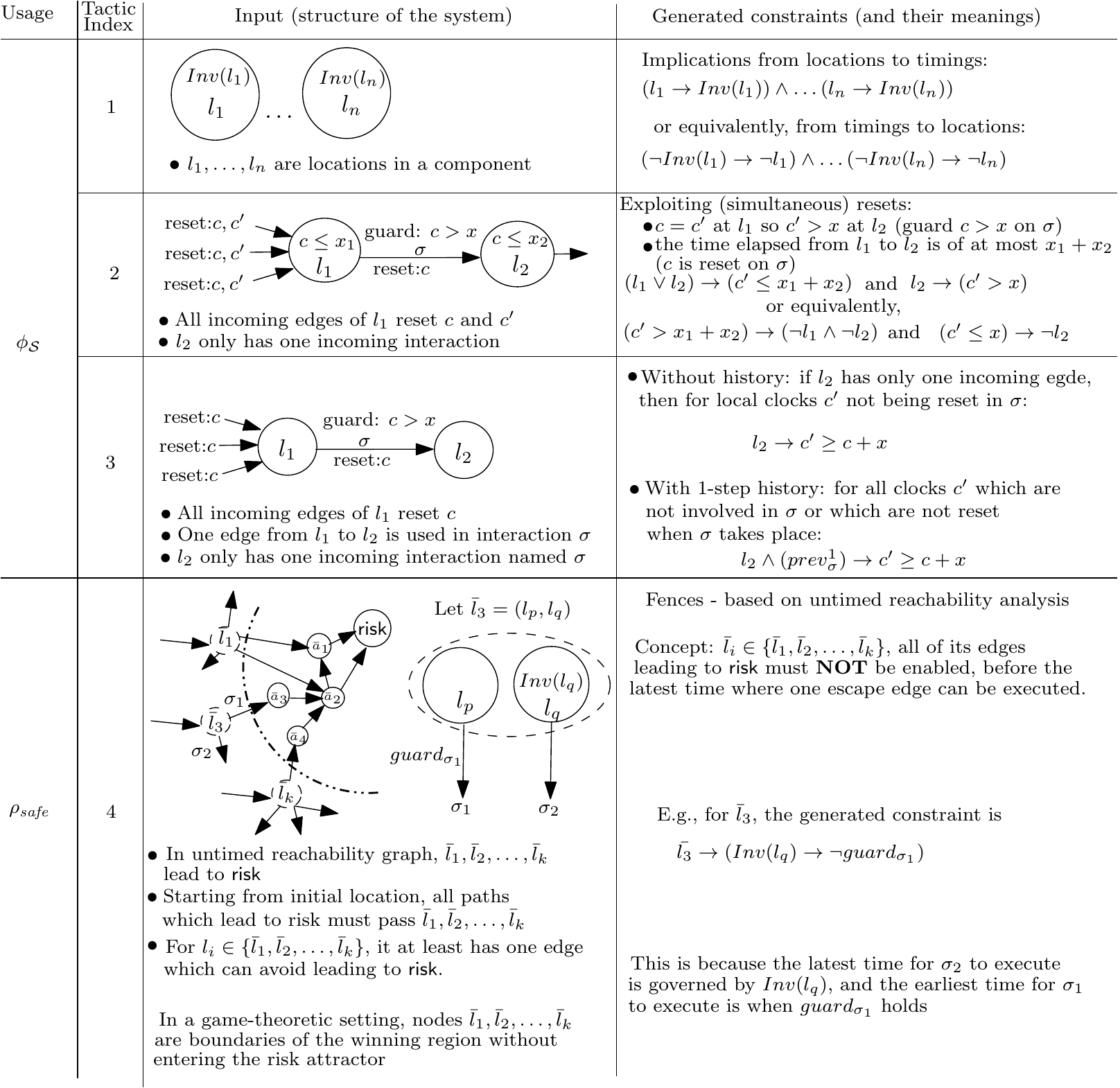}
	\caption{A list of tactics for generating system invariants  (tactic~1 to~3) and $\rho_{\mathit{safe}}(x,y)$ (tactic~4). 
}
	\label{fig:static.analysis}
\end{figure}



\subsection{Generating $\exists\forall$-Constraints for Safety Synthesis}\label{sub.sec.parameterized.invariants}
\comments{
	\obs{wrt Fig 3, no need to put equivalent forms... one should be
		enough. I'd rather use $\bar{l}$ instead of $n_i$ (to denote locs in
		Abs) because the notation would be more uniform this way. and
		$\sigma_i$ from $l_i$ is confusing as $\sigma$ are interactions and
		$l_i$ are locs in components, so i'd expect an action instead. I'd
		also keep just one $\isyst$ instead of 3. I can modify the fig, if i
		can have the source code.}
	}

Now, we reduce timed orchestration synthesis problem for safety properties to corresponding $\exists\forall$SMT constraints of the form
\begin{align}
\exists x \in \phi_C: \, \forall y: \isyst(x, y) \rightarrow (\phi_{\mathit{state}} \wedge \neg \phi_{deadlock}(x,y)\wedge \rho_{\mathit{safe}}(x,y))\mbox{,}
\label{eq:synt}
\end{align}
where $x$ is the set of unknown variables to be synthesized, 
$y$ is the set of clocks, locations, 
optional variables for encoding the history of interactions, 
$\isyst(x, y)$ is the
summary as an over-approximation of system dynamics, 
$\neg \phi_{deadlock}(x,y)$
is the translation of $\pintt$ into a safety property as described in
Section~\ref{sub.sec.monitor}, and $\rho_{\mathit{safe}}(x,y)$ is a
disjunction of sufficient conditions for not reaching location
\sig{risk}\@.
 Constraints $\phi_C$ and $\phi_{\mathit{state}}$ are given system requirements
as mentioned in the problem formulation in
Section~\ref{sec.background}\@.  For ease of reading, we use the
notation $k > 0$, respectively $k = 0$, to distinguish the case when
$k$-step interaction-history is encoded by means of universal
variables from the case when interaction-history is not used at all in
the generation of $\isyst(x, y)$. We note that any solution for
Formula~\eqref{eq:synt} is a solution to the problem formulated in
Section~\ref{sec.background}.


\subsubsection{Generating $\isyst(x, y)^{k=0}$.} 

Our approach to characterize the behavior of the system is
\textit{compositional}. This way, we avoid computing the whole
product, which is, in most non-trivial cases, a costly
operation. Instead, our computed invariant is the conjunction of the
following three: (1) invariants for each component, (2) invariants
capturing conditions when synchronization appears, and (3) untimed
reachability.
\begin{enumerate}[label=(\arabic*)]
	\item\label{it:ci} \emph{Component invariants} $\ci(C_i)$ are
          properties characterizing components $C_i$. We do not
          restrict their computation to a specific methodology. What
          matters is that such properties can be shown to be
          invariants. In our framework, where components are
          parametric timed automata, one way to obtain invariants is
          to compute abstractions\footnote{In general, the
            reachability problem is undecidable \cite{alur93}. We
            refer to~\cite{jovanovic13:synthesis-pta} as a pointer for the
            computation of symbolic state abstractions.} of classical
          zone graphs~\cite{henzinger94}. Zone graphs are symbolic
          representations of the reachable state space of parametric
          timed automata. In practice, easier solutions work as
          well. One example is the tactic~1 in
          Figure~\ref{fig:static.analysis}. As an illustration, for
          $\phio_1$, by applying tactic~1, the resulting invariant is
          $(p_{10} \rightarrow t_{s_1}\leq \alpha_1) \wedge (p_{11}
          \rightarrow t_{1}\leq \gamma_1) \wedge (p_{12} \wedge
          t_{1}\leq 3) \wedge ( p_{13} \rightarrow t_{1}\leq 3)$. By
          applying tactic~2 and~3 one can derive the additional
          conditions for $\phio_1$ and $\phio_2$:
   
    \begin{itemize}
    	\item $(ts_1 < \eta_1) \rightarrow (p_{11})$ and
    	$(ts_2 < \eta_2) \rightarrow (p_{21})$.
    	\item $(ts_1 < \eta_1 + 2) \rightarrow (\neg p_{10}
    	\wedge \neg p_{13})$ and $(ts_2 < \eta_2 + 3)
    	\rightarrow (\neg p_{20} \wedge \neg p_{23})$.
    	\item $(ts_1 > \gamma_1 + 3) \rightarrow (\neg p_{11}
    	\wedge \neg p_{12})$ and $(ts_2 > \gamma_2 + 4)
    	\rightarrow (\neg p_{21} \wedge \neg p_{22})$.
    	\item $(ts_1 > \gamma_1 + 3 + 3) \rightarrow (p_{10})$
    	and $(ts_2 > \gamma_2 + 4 + 2) \rightarrow
    	(p_{20})$.
    \end{itemize}

	\item\label{it:ii} \emph{Discrete-jump invariants} $\varPi_{\mathcal{S}}$
          are global clock constraints inferred either (a) statically
          from resets on incoming transitions or (b) from the
          simultaneity of interactions and the synchrony of time
          progress. Such constraints are generated by applying
          tactics~2 and~3 of Figure~\ref{fig:static.analysis}.
	\begin{enumerate}[label=(\alph*)]
		\item Consider location $p_{12}$ in $\phio_1$. It has
                  one incoming edge which resets clock $t_1$. As no
                  other clock in the system is reset, and the incoming
                  edge has guard $t_1 \geq \eta_1$, one derive that $(
                  p_{12}) \rightarrow (t_{s_1}-t_1 \geq \eta_1)$,
                  i.e., in location $p_{12}$, all other local clock
                  readings should at least be $\eta_1$ unit larger
                  than $t_1$.
		
		\item Consider interaction \sig{reset}. It leads to
                  location $(p_{11}, -, -, p_{21})$, where initial
                  state is located. One can derive the invariant
                  $(p_{11} \wedge p_{21}) \rightarrow
                  (t_{s_1}=t_1=t_{s_2}=t_2)$.  By a similar argument,
                  one can also infer that $t_{s_1}=t_{s_2}$ holds, due
                  to unique clock reset action on \sig{reset}.
	\end{enumerate}

	\item \emph{Untimed abstract reachability invariant}
          $Abs(\mathcal{S})$ is the set of reachable location
          combinations of $\mathcal{S}$ by ignoring clocks and by only
          considering the lockings by interactions. E.g., with untimed
          reachability analysis from initial locations, one can deduce
          that $(p_{11} \wedge p_{21}) \rightarrow (f_{10} \wedge
          f_{20})$, i.e., buffers are not occupied before both robots
          start. Notice that $Abs(\mathcal{S})$ is not sensitive to
          parameter change due to its ignoring of clocks.

	\end{enumerate}


\begin{remark}
Commonly, a tactic creates constraints of the form $\phi_{loc}
\rightarrow \phi_{clock}$, where $\phi_{loc}$ is a formula over
locations and $\phi_{clock}$ is a property associated with clocks.  As
$\phi_{loc} \rightarrow \phi_{clock} \equiv \neg \phi_{clock}
\rightarrow \neg \phi_{loc}$, the $\exists\forall$-solver also uses
such constraints to reason that under concrete timing conditions, it
is impossible to be in a state in $Abs(\mathcal{S})$.
To illustrate this, we return to the robot example. In the
untimed setup, $\phio_1$ and $\phio_2$ can execute \port{take1l} and
subsequently \port{take2l}. Therefore, state $(p_{12}, p_{22},
\sig{risk})$ is within $Abs(\mathcal{S}_{inv,k})$. However, under a
parameter assigment as $\gamma_1=\eta_1:=0$ and $\gamma_2=\eta_2:=15$,
the constraint solver invalidates such a state by the following
reasoning:
	
	\begin{itemize}
		\item When $\phio_2$ is at $p_{22}$, $t_{s_2} \geq
		\eta_2$, i.e., $t_{s_2} \geq 15$ (by tactic~2 in
		Figure~\ref{fig:static.analysis}).
		\item $t_{s_2}=t_{s_1}$ (from Item~\ref{it:ii}),
		so $t_{s_1} \geq 15$.
		\item As $(t_{s_1} > \gamma_1 + 3 + 3) \rightarrow
		(p_{10})$ (from  Item~\ref{it:ci}) and
		$\gamma_1 = 0$, $\phio_1$ must stay in
		$p_{10}$.

	\end{itemize}
		Therefore, the reachability of $(p_{12}, p_{22}, \sig{risk})$
		in $Abs(\mathcal{S}_{inv,k})$ is invalidated under parameter assignment
		$\gamma_1=\eta_1:=0$ and $\gamma_2=\eta_2:=15$.
\end{remark}
	
We define $\isyst(x, y)^{k=0}$ as $\bigwedge_{C_i \in C} \ci(C_i) \wedge  \varPi_{\mathcal{S}} \wedge Abs(\mathcal{S})$ and denote $\isyst(x, y)^{k=0}(\vec{v})$
to be the result of replacing the unknown variables $V$ by assignment
$\vec{v}$ in $\ci(C_i)$ and $\varPi_{\mathcal{S}}$. Using the fact
that the conjunction of invariants is an invariant itself, it can be
shown that indeed $\isyst(x, y)^{k=0}$ is an invariant of $\syst$.

\begin{lemma}
\label{lem:inv}
	For any assignment $\vec{v}$ for unknown parameters,
        $\isyst(x, y)^{k=0}(\vec{v})$ is an invariant of
        $\mathcal{S}(\vec{v})$.
\end{lemma}

\subsubsection{Generating $\isyst(x, y)^{k>0}$.} 
For $\exists\forall$-constraint solving, the precision of system
invariants plays an important role. 
Given the set of interactions $\Sigma$, one
can introduce a set of Boolean variables $\{prev_{\sigma}^{k}\,|\, \sigma\in \Sigma\}$ to record $k$-previously executed 
interaction. As an example, consider tactic~2 in
Figure~\ref{fig:static.analysis}. When one records the previously
executed interactions, the condition $c' \geq c+x$ is associated with
location $l_2$ and the previously executed interaction $\sigma$.
Assume that $l_2$ has another incoming interaction $\sigma'$, which
does not reset $c$. Then a memoryless approach (i.e., no history)
needs to take the disjunction of conditions from all incoming edges,
thereby losing the knowledge of $c' \geq c+x$. 


The price for recording $k$-step interaction history, given $\Sigma$
as the set of interactions, is only at the cost of introducing
$k|\Sigma|$ Boolean variables as universal variables. In a similar manner as for Lemma~\ref{lem:inv}, it can be shown that
$\isyst(x, y)^{k>0}$ is an invariant of the system.

\subsubsection{Generating $\rho_{\mathit{safe}}(x,y)$ with ``fence'' constraints.} 

An intuitive yet sometimes sufficient way is to assign
$\rho_{\mathit{safe}}(x,y)$ to be simply $\neg \sig{risk}$. However, one can
also introduce other constraints $\rho_1, \ldots, \rho_n$, where each
of them is a sufficient condition to block the run to enter
$\sig{risk}$, and set $\rho_{\mathit{safe}}(x,y) := (\neg \sig{risk}) \wedge
\bigvee_{i=1}^n \rho_i$, and leave the finding of solutions to the
$\exists\forall$-solver. The computation of these constraints should
be light-weight. Here we present the \emph{fence-condition tactic}
(index~4 of Figure~\ref{fig:static.analysis}) which only involves the
computation of backward untimed reachability and the static scan of
components.

The underlying concept is to find a set of nodes $\bar{l}_1,
\bar{l}_2, \ldots, \bar{l}_k$ in the abstract reachability graph,
where every path that leads to $\sig{risk}$ must pass one node
$\bar{l}_i \in \{\bar{l}_1, \bar{l}_2, \ldots, \bar{l}_k\}$. At each
node $\bar{l}_i$, there exists at least an ``escape edge'' which can
avoid leading to risk. Finding such a set is done by solving a safety
game (using standard attractor computation defined in two-player,
turn-based games over finite arena; see~\cite{attractor} for
details) with all nodes viewed as control vertices. Here we explain
the attractor concept using examples. In
Figure~\ref{fig:static.analysis}, the computation of attractors adds
gradually $\{\bar{a}_1\}$, $\{\bar{a}_2\}$ (as one outgoing edge leads
to \sig{risk} and the other leads to $\bar{a}_1$), $\{\bar{a}_3,
\bar{a}_4\}$ to the attractor of $\sig{risk}$. Nodes such as
$\bar{l}_3$ are outside the attractor, as it can use $\sigma_2$ to
escape.

With $\{\bar{l}_1, \bar{l}_2, \ldots, \bar{l}_k\}$ identified, whenever one can guarantee
that at node $\bar{l}_i$, interactions which leads to the attractor will
never be executed, then one can guarantee that $\sig{risk}$ is never
reached from the initial state for any time run. For $\bar{l}_i \in \{\bar{l}_1, \bar{l}_2, \ldots, \bar{l}_k\}$, let $\Sigma_{attr,i}$ be outgoing interactions which
leads to attractor and $\Sigma_{str,i}$ be the winning strategy on
$\bar{l}_i$ to escape from the attractor. For interaction $\sigma$, let
$guard_{\sigma}$ be the guard condition for which $\sigma$ can take
place. We restrict ourselves to such that guards are conjunctions of
form $clock \sim k$ where $\sim\; \in\{>,\geq\}$.  Then we can create
the following constraint:

\[\bigwedge_{i=1}^{k} \bigwedge_{\sigma \in \Sigma_{attr,i}} (\bar{l}_i \rightarrow \bigvee_{\sigma' \in \Sigma_{str,i}} (\tEn{\sigma'} \rightarrow \neg guard_{\sigma} ))\]

Intuitively, the constraint specifies that at $\bar{l}_i$, as long as when
an interaction $\sigma'$ from $\Sigma_{str,i}$ can be executed in the
future (i.e., $\tEn{\sigma'}$), interaction $\sigma$ in
$\Sigma_{attr,i}$ should not be enabled. In
Figure~\ref{fig:static.analysis}, for node $\bar{l}_3$, as $\tEn{\sigma_2}$
is merely the invariance condition on $l_q$, we have $\bar{l}_3 \rightarrow
(Inv(l_q) \rightarrow \neg guard_{\sigma_1})$.



\subsection{Finding Satisfying Instances for $\exists\forall$ Formulas}
\label{sub.sec.efsmt}

\comments{
	\begin{figure}[t]
		\centering
		\includegraphics[width=0.7\columnwidth]{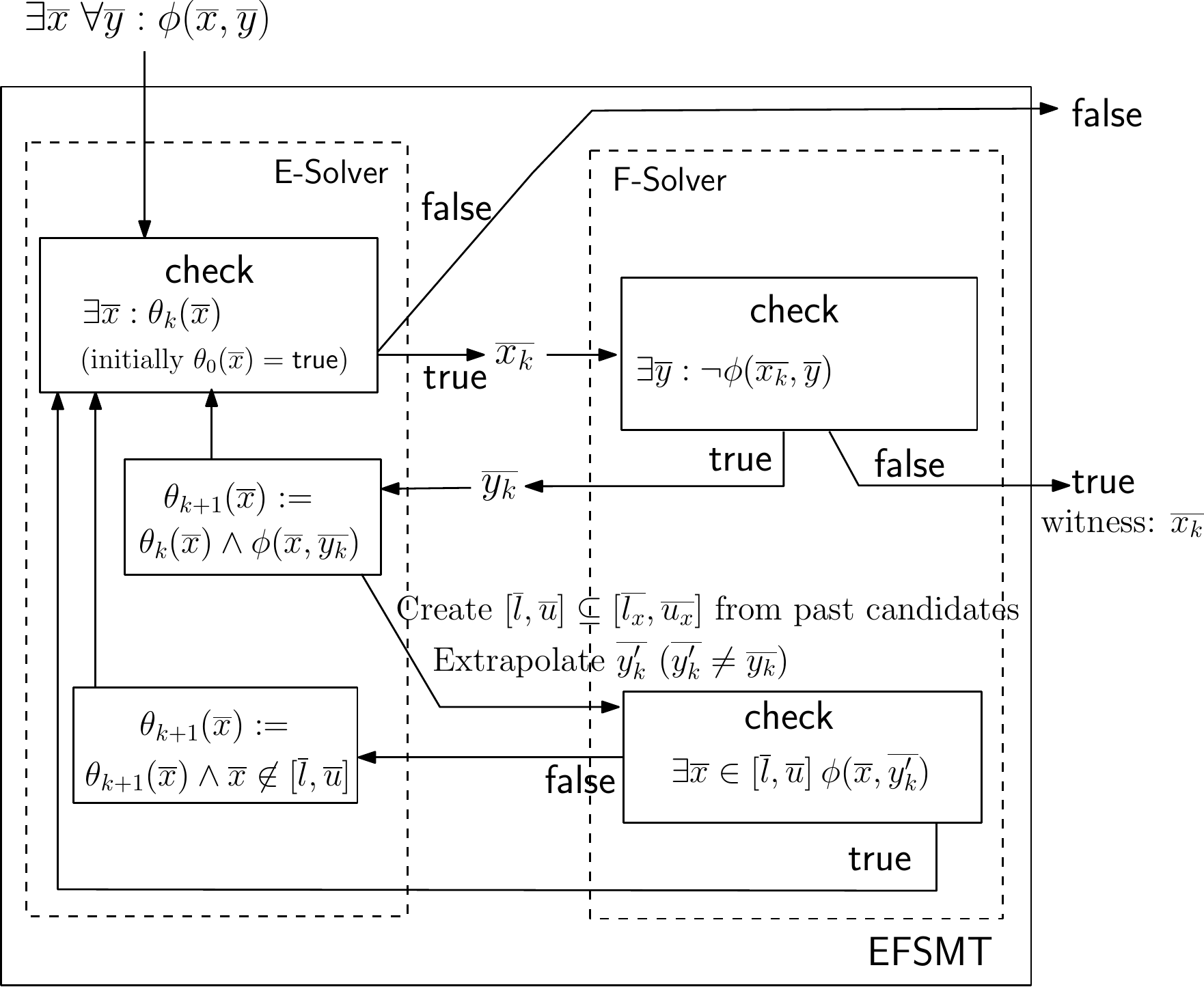}
		\caption{Algorithmic flow for \textsf{EFSMT}.}
		\label{fig:algorithm}
	\end{figure}
}



We outline a verification procedure implemented in \efsmt for solving
constraint problems of the form $\exists x\, \forall y:\phi(x, y)$,
where $\phi(x, y)$ is a quantifier-free formula involving two variable
sets $x$ and $y$. This class is generic enough to fit formulas such as
Formula~\eqref{eq:synt} in
Section~\ref{sub.sec.parameterized.invariants}. The verification
procedure is based on two SMT solver instances, the so-called \esolver
and \fsolver. These two solvers are applied to quantifier-free
formulas of different polarities in order to reflect the quantifier
alternation, and they are combined by means of a \emph{counter-example
  guided refinement} strategy.


At the $k$-th iteration, the \esolver either generates an
instance $x_k$ for $x$ or the procedure returns
with \textsf{false}\@.  An $x_k$ provided by the
\esolver is passed to the \fsolver for checking if
$\exists y:\neg\phi(x_k, y)$ holds.
If not, then $x_k$ is the witness for the problem $\exists x \forall y:\phi(x, y)$.
In case there is a satisfying assignment $y_k$ generated
 at the $k$-th iteration, the \fsolver passes the
constraint $\phi(x, y_k)$ to the
\esolver, for ruling out such $x$ as potential
witnesses.  Future candidate $x_{k+1}$ from the $k+1$-th
iteration should therefore not only satisfy $\phi(x_{k+1},
y_0), \ldots, \phi(x_{k+1}, y_{k-1})$
but also allow $\phi(x_{k+1}, y_k)$ returning
\textsf{true}\@ (an example is listed below, for the ease of understanding).


In many cases the domain of integer parameters is bounded, and the \efsmt solving algorithm is terminating\footnote{In general, the pure usage of two
quantifier-free solvers do not guarantee termination~\cite{DBLP:journals/corr/ChengSRB13}.}\@, as there are only finitely many variable assignments. 
Consider, for example, a  constraint such as $\exists x_1 \in [0, 100]\cap
\mathbb{Z}, \forall y_1 \in [10, 20] \cap \mathbb{R}: x_1 - y_1 \geq
80$. Assume that the explicit enumeration and \efsmt both start with
the order $x_1 = 0, 1, \ldots, 100$. In these cases,  brute-force
enumeration method need to iterate 100 times, until they find $x_1=100$
(the only satisfying instance). For \efsmt, if $x_1 = 0$, then the
counterexample provided by \fsolver, for instance, $y_1=10$, falsifies
it. After this step, \esolver creates a new assignment by ensuring $x_1
- 10 \geq 80$ thus immediately jumping to $x_1 = 90$. Consequently, it
omits checking assignments $x_1 = 1, \ldots, 89$. In other words, our solver
may be viewed  as an {\em acceleration of explicit enumeration} of SMT via counterexamples.

\section{Extensions}



Due to the reduction of timed orchestration problems to $\exists\forall$SMT on can readily handle richer arithmetic constraints in synthesis problems. 
We briefly outline how quantitative synthesis, robustness synthesis, and synthesis beyond PTA may be encoded.

\noindent\textbf{Quantitative Synthesis.} In practice one is usually interested in obtaining parameters for optimized system behavior  (e.g., \sig{min}, lexicographic)\@.
For example, one might be interested in obtaining a minimum value for the parameter $\alpha_1$ in our running example in in Figure~\ref{fig:TimedPhilosopher}\@. 
In solving the corresponding $\exists\forall$SMT constraints using the proposed two solver approach, one may simply use an \esolver with optimization capabilities --- e.g. a MaxSMT solver such as $\nu$Z~\cite{BjornerPF15}) --- instead of an SMT solver. 
In this way, the proposed solution of the \esolver is optimal with respect to the current set of constraints.



\noindent\textbf{Robustness Synthesis.} Using  $\exists\forall$SMT constraints, the imprecision of system may be modeled by means of universally-quantified, bounded variables.
For example, one may model the imprecision for a a guard $t_1>2$ by $t_1>2+\delta$, where $\delta \in[-0.05, 0.05]$\@, and $\delta \in[-0.05, 0.05]$ is added as a new
universally-quantified variable in the  $\exists\forall$SMT\@. 


\noindent\textbf{Beyond PTA.} Using the full expressivness of $\exists\forall$-constraints, one may also encode guards, for instance, $t_1 + 3\,t_4 \geq 10$, which go beyond clock 
constraints of plain PTAs. 




\section{Evaluation}\label{sec.evaluation}

The above extensions come for free with our prototype
tool\footnote{\url{http://www.chihhongcheng.info/efsmt}} which we have
developed for implementing the concepts in
Section~\ref{sec.workflow}. Technically, the prototype automatically
generates monitor components based upon the LTL2Buchi
transformation~\cite{Giannakopoulou02fromstates} for generating
B\"uchi automata. The symbolic reachability underlying the computation
of $Abs(\mathcal{S})$ and the attractor computation for fence conditions use
JDD\footnote{\url{http://javaddlib.sourceforge.net/jdd/}}, a Java
package for efficiently manipulating Binary Decision Diagrams
(BDDs). The construction of the $\exists\forall$ constraint solver is
based upon the combination of our \textsf{E-solver} and
\textsf{F-solver} which in turn wrap SMT-solver
Yices2~\cite{dutertre2014yices} when quantifier-free constraint
solving is needed. 

\begin{table}[t]
	\scriptsize
	\centering
	\begin{tabular}{|c|c|c|c|c|c|}
		\hline
		Example	 & num para. ($\exists$) & 
		parameter range & num $\forall$ variables   & \textsf{EFSMT} time (sec) \\ \hline
		\phio3		 & 12 & $[0,30]\cap \mathbb{Z}$ & 6 real, 9 Bool  & 1.262 \\ \hline
		
		\phio4	 & 16 & $[0,30]\cap \mathbb{Z}$ & 8 real, 12 Bool   & 3.037 \\ \hline
		
		\phio5	 & 20 & $[0,30]\cap \mathbb{Z}$ & 10 real, 15 Bool  & 33.424 \\ \hline
		\phio6	 & 24 & $[0,30]\cap \mathbb{Z}$ & 12 real, 18 Bool & 165.856 \\ \hline
		\phio7	 & 28 & $[0,30]\cap \mathbb{Z}$ & 14 real, 21 Bool  & from 154.958 to 1026.992\\ 	\hline
		Worker 10	 & 1 & $\beta \leq 1000$ & 10 real, 12 Bool & 0.040 \\ \hline
		Worker 20	 & 1 & $\beta \leq 1000$ & 20 real, 22 Bool & 0.079 \\ \hline
		Worker 30	 & 1 & $\beta \leq 1000$ & 30 real, 32 Bool & 0.195 \\ \hline
		Worker 40	 & 1 & $\beta \leq 1000$ & 40 real, 42 Bool & 0.371 \\ \hline
		Worker 50	 & 1 & $\beta \leq 1000$ & 50 real, 52 Bool & 0.561 \\
		
		\hline
	\end{tabular}
	
	\vspace{2mm}
	
	\begin{tabular}{|c|c|c|c|c|}
		\hline
		Example & num para. ($\exists$) & 
		parameter range & num $\forall$ variables  & \textsf{IMITATOR} time (sec) \\ \hline
		\phio5		 & 20 & $[0,30]\cap \mathbb{Z}$ & 10 real, 15 Bool & 570.88 \\ \hline
		\phio6		 & 24 & $[0,30]\cap \mathbb{Z}$ & 12 real, 18 Bool & t.o. ($>3600 sec$) \\ \hline
		\phio7		 & 28 & $[0,30]\cap \mathbb{Z}$ & 14 real, 21 Bool & t.o. ($>3600 sec$)\\ \hline
	\end{tabular}
	
	\vspace{2mm}
	
	\begin{tabular}{|c|c|c|c|c|}
		\hline
		Example & num para. ($\exists$) & 
		parameter range & num $\forall$ variables  & \textsf{UPPAAL} ``verification" time (sec) \\ \hline
		\phio7	 & 28 & $[0,30]\cap \mathbb{Z}$ & 14 real, 21 Bool  & 15.56\\ 	\hline
		Worker 50	 & 1 & $\beta \leq 1000$ & 50 real, 52 Bool & t.o. ($>600 sec$) \\ \hline
	\end{tabular}
	\vspace{2mm}
	\caption{Evaluation results for deadlock, and comparison with other synthesis and verification tools. Worker is a modified (by creating unknowns) example from~\cite{AstefanoaeiRBBC14}.}
		\label{experiment.result.state-based.properties}

\end{table}

\begin{table}[t]
	\scriptsize
	\centering
	\begin{tabular}{|c|c|c|c|c|c|}
		\hline
		Example	& Property &  num.  $\exists$ & 
		parameter range & num $\forall$ variables  & \textsf{EFSMT} (sec) \\ \hline
		\phio3		& Deadlock-free, LTL & 12 & $[0,30]\cap \mathbb{Z}$ & 6 real, 12 Bool & 1.412 \\ \hline
		\phio4		& Deadlock-free, LTL & 16 & $[0,30]\cap \mathbb{Z}$ & 8 real, 15 Bool & 17.765 \\ \hline
		\phio5		& Deadlock-free, LTL  & 20 & $[0,30]\cap \mathbb{Z}$ & 10 real, 18 Bool & 301.665 \\ \hline
		\phio6		& Deadlock-free, LTL & 24 & $[0,30]\cap \mathbb{Z}$ & 12 real, 21 Bool  & 4262.047 \\ \hline
		 MES2		& Error handling (using \U) & 2 & $[0,100]\cap \mathbb{Z}$ & 2 real, 12 Bool & 0.041 \\ \hline
		 MES2		& Error handling (using \U)  & 2 & $[0,40]\cap \mathbb{Z}$ & 2 real, 12 Bool & 0.169 (no solution) \\ \hline
		  MES2		& Parameterized handling (using \U) & 2 & $[0,200]\cap \mathbb{Z}$ & 2 real, 2 int, 12 Bool & 0.135  \\ \hline
		 MES2		& Parameterized handling (using \U) & 2 & $[0,100]\cap \mathbb{Z}$ & 2 real, 2 int, 12 Bool & 0.479 (no solution) \\ \hline
			 MES3 	& Prod. seq. control 1 (using \textbf{X}, \F)  & 4 & $[0,100]\cap \mathbb{Z}$ & 3 real, 18 Bool & 0.204 \\ \hline
		 MES3 	& Prod. seq. control 2 (using \textbf{X}, \F) & 4 & $[0,100]\cap \mathbb{Z}$ & 3 real, 18 Bool & 5.341 (no solution) \\ \hline
	
	\end{tabular}
	\vspace{2mm}
	\caption{Experimental results for LTL properties.}
	\label{experiment.result.ltl}

\end{table}

Tables~\ref{experiment.result.state-based.properties}
and~\ref{experiment.result.ltl} show the results of our initial
evaluation (under Intel i5-4300u CPU, 8GB RAM, Ubuntu 14.04 64-bit
OS). The recorded execution times for other tools (e.g., IMITATOR) are
based on the newest tool versions available for download. For the
robot problem in Table~\ref{experiment.result.state-based.properties},
the constants in one automaton differ from those in the other
automaton. This is in order to avoid symmetric effect and more
importantly and additionally, to be closer to more realistic settings.
As an example, using the same experiment setup to run IMITATOR for
five robots already takes about ten minutes (\textsf{EFSMT} is about
one order of magnitude faster).  In
Table~\ref{experiment.result.state-based.properties}, we do not list
the time needed for generating constraints, as it is neglectable
compared to $\exists\forall$-constraint solving (for abstract
reachability, even for 10 robots it takes less than 5
seconds). However,  the ordering of the constraints
may greatly influence timings. Consequently, obtaining good
results for synthesizing parameters to enforce safety properties
requires both a good solver and a tailored constraint structure
suitable for exploiting the locality of constraints. In our case, this
is truly possible thanks to our local component invariants.

From Table~\ref{experiment.result.state-based.properties} readers may
be surprised by the timing for ensuring promptness in the case of 6
robots. The increase in computation time follows from \textsf{EFSMT}
searching for all possibilities without finding any, as the sum of all
mode upper bounds is greater than~30. Another interesting behavior
which occurred during our evaluation exhibits that the non-determinism
within SMT solvers (for $\phio_7$ it creates multiple satisfying
assignments) may drastically influence performance.

Table~\ref{experiment.result.state-based.properties} also shows the
result of analyzing the temperature controller problem modified
from~\cite{AstefanoaeiRBBC14}, where only one unknown parameter needs
to be synthesized.  In the experimental setup, the search starts from
$\beta = 999$, then it quickly prunes the search space and identifies
the result in about 2 to 5 steps. This is the reason why the
computation time is surprisingly small, and clearly demonstrates the
superiority of \sig{EFSMT} over a brute-force enumeration
method. However, as our parameterized timer invariant generation is
far from precise, our generated result is not optimal. Still, for
verifying our result using UPPAAL, it takes more than 10 minutes for
50 workers. 
This demonstrates that at least some problems may be solved by
inferring synchronization properties without paying the price of doing holistic state space exploration. 


\subsection{Flexible Production System Case Study}\label{product.parameter}
	
\begin{figure}[t]
	\centering
	\includegraphics[width=0.7\columnwidth]{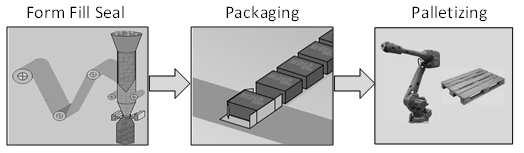}
	\caption{A sample cereal packaging line.}
	\label{fig:food.packaging.line}
\end{figure}

In discrete manufacturing, individual workpieces are treated in multiple processing steps, 
typically organized sequentially with multiple machines. Under the initiatives of {\em Industrie 4.0}, it is
generally perceived that machines can communicate their status, mainly
on their state changes. This view fits well with our methodology. To
see this, it suffices to adopt the interpretation where one can
isolate the functionality of every machine as components with
parameters and design each component without the use of global
clocks. Along these lines, as an application of our method to discrete
manufacturing, we use simplified packaging line as a case study
in the food \& beverage segment. The main components are displayed in
Figure~\ref{fig:food.packaging.line}\@. More precisely,
Figure~\ref{fig:food.packaging.line} illustrates a Form-Fill-Seal
(FFS) machine which fills parts produced in the upstream process into
plastic bags. In turn, the plastic bags are packaged into boxes by a
packaging machine. Finally, cartons are placed on a pallet for
shipment. We assumed that the product to be created is breakfast
cereal, while retailers can request variations on bag size and box
capacity in terms of $x$ grams per bag and $y$ bags in one box. To
handle such product variations realized by the two variables $x$ and
$y$, we simply need to encode them as universal variables. On the
other hand, FFS machine parameters are encoded as existential variables:
the execution times for filling, respectively sealing, are configured
by $\alpha$, respect($\beta$ sec).  In the automaton for FFS, these
variables are placed as the guards and location conditions to
represent the lapse of time. By encoding the problem into \sig{EFSMT},
we are able to synthesize $\alpha$ and $\beta$ such that it works for
all $x$ and $y$ specified in the range. For example, a typical
encoding is $\exists \alpha, \beta \,\forall x \in [100, 300] \cap
\mathbb{R}\, \forall y \in [10, 24] \cap \mathbb{Z}$.

For this scenario we formulate a system description together with properties 
for excluding undesired action sequences such as
``when the packaging station buffer is full, FFS should stop shipping
until the buffer has space''\@. This property is encoded in terms of the interaction-level LTL formula 
   $$\G(\sig{Packaging.stackfull} \rightarrow (\neg
\sig{FFS.ship}\; \U\; \sig{Packaging.stackavailable}))\mbox{.}$$

\smallskip

\noindent\textbf{Applicability and limitations.}  We apply our
solver for solving the timed orchestration problem on interaction-level properties;
the results of this case study are summarized in Figure~\ref{experiment.result.ltl}\@. 
For those properties where \efsmt successfully synthesizes parameters, 
we also tried to restrict the domain and recorded required time for \efsmt to report
``unable to find a solution''\@. 
Our solving approach seems to scale well because of the use of compositional techniques, 
but at the expense of precision for relations between clocks from different components.  
Moreover, due to recording the history of
interactions, our solver seems to perform well on LTL
formulas include $\F$ or $\U$, since these
properties are translated into a template  ``whenever an event
occurs, something good should happen within a finite number of steps''
by means of unrolling. 
Finally, we note that
constraint grouping and variable ordering plays an important role in
the performance of the underlying SMT solver Yices~2\@. 
More precisely, we observe in our experiments a sever performance penalty
whenever constraints are not properly grouped or whenever the evaluation order of
variables does not respect the grouping. 
Informally,  a constraint grouping may be called {\em proper} if the grouping in \textsf{EFSMT}
follows that of the constraints in the invariants for untimed reachability. 
These invariants are computed by means of BDDs and FORCE
ordering heuristics~\cite{aloul2003force} in our implementation, which
results in relatively compact representations and to 
also reduce the size of invariant constraints.

\section{Conclusions} \label{sec.summary} 

The main contributions of this paper include (1) encoding of line integration problems in terms of timed orchestration synthesis,
(2) upper bound on the number of unrolling steps in bounded synthesis for PTA, 
(3) encoding of timed orchestration synthesis in terms of $\exists\forall$SMT,  and
(4) set of computationally-cheap over-approximations for avoiding overly eager and expensive computations of the precise parametric images of the set of reachable states\@.
Some of  the key ingredients of this logical approach to solving timed orchestration problems include the
translation of deterministic monitors from LTL properties, the generation of parametric invariants, the use of two SMT solvers for
$\exists\forall$ constraints, constraint grouping and variable ordering. 
We demonstrate the feasibility of this approach by means of solving some typical line integration problems as encountered in industrial practice;
it still remains to be seen, however, if and how the proposed methodology and tools scales to solving orchestration problems 
for real-world production lines.
In future work, we therefore plan to go beyond $\omega$-words when considering interaction-level LTL properties, develop static analysis techniques on the 
system structure for obtaining cheap invariants, investigate hierarchical solving approaches, and to extend the orchestration synthesis problem
to hybrid systems.









\mycomment{
\appendix
\section{Appendix}
Readers are welcome to try how generated invariants are executed under the new version of $\textsf{EFSMT}$ (requires installation of Yices2) by accessing the following website: 
$$\texttt{http://www.chihhongcheng.info/efsmt}$$
}

\end{document}